\documentclass[prx,aps,showpacs,twocolumn,preprintnumbers,amsmath,amssymb,superscriptaddress,longbibliography]{revtex4-2}
\usepackage[english]{babel}
\usepackage{amsmath,amssymb,bbm,mathrsfs,mathtools,multirow,diagbox,xcolor,%
  graphicx,tabularx,comment,amsfonts,dsfont,lettrine,units,comment}

\usepackage{graphicx}
\usepackage[colorlinks=True,linkcolor=red,citecolor=blue,urlcolor=blue]{hyperref}
\usepackage{bookmark}
\usepackage{xcolor}
\usepackage{cases}
\usepackage{chngcntr}
\usepackage{braket}
\usepackage{nicefrac}
\usepackage{bm}
\usepackage{bbm}
\usepackage{braket}
\usepackage{lineno}
\usepackage{array}
\newcolumntype{C}{>{$}c<{$}}
\usepackage{newtxtext} 
\usepackage{orcidlink}
\usepackage[normalem]{ulem} 
\usepackage{wasysym} 
\usepackage{cancel}

\newcommand{\bs}{\boldsymbol}

\usepackage{graphicx}
\usepackage{bm}

\usepackage{hyperref}
\hypersetup{
    colorlinks=true,
    linkcolor=blue,
    filecolor=blue,      
    urlcolor=blue,
    citecolor=blue,
    pdftitle={localizer},    
    pdfauthor={Adolfo G. Grushin},     
    pdfkeywords={localizer}, 
}
 
\makeatletter
\renewcommand*\env@matrix[1][*\c@MaxMatrixCols c]{%
  \hskip -\arraycolsep
  \let\@ifnextchar\new@ifnextchar
  \array{#1}}
\makeatother

\begin{document}

\title{
{Topological zero-modes of the spectral localizer of trivial metals}
}

\author{Selma Franca\orcidlink{0000-0002-0584-2202}}
\email{selma.franca@neel.cnrs.fr}
\author{Adolfo G. Grushin\orcidlink{0000-0001-7678-7100}}
\email{adolfo.grushin@neel.cnrs.fr}
\affiliation{Univ. Grenoble Alpes, CNRS, Grenoble INP, 
Institut Néel, 38000 Grenoble, France}

\date{\today}

\begin{abstract}

{Topological insulators are described by topological invariants that can be computed by integrals over momentum space, 
but also as traces over local, real space topological markers.
These markers are useful to detect topological insulating phases in disordered crystals, quasicrystals and amorphous systems.
Among these markers, only the spectral localizer operator can be used to distinguish topological metals, that show zero-modes of the localizer spectrum.
However, it remains unclear whether trivial metals also display zero-modes, and if their localizer spectrum is distinguishable from topological ones.
Here, we show that trivial metals generically display zero-modes of the  
localizer spectrum.
The localizer zero-modes are determined by the zero-mode solutions of a Dirac equation with a varying mass parameter.}
We use this observation, valid in any dimension, to determine the difference between the localizer spectrum of trivial and topological metals,  
and conjecture the spectrum of the localizer for fractional quantum Hall edges.
Because the localizer is a local, real space operator, it may be used as a tool to differentiate between non-crystalline topological and trivial metals,  
and characterize strongly correlated systems, for which local topological markers are scarce.

\end{abstract}

\maketitle

\section{Introduction}

{
Topological phases of matter
are identified by non-zero bulk topological invariants that guarantee their protection against disorder~\cite{hasan2010, Qi_Zhang2011, Armitage2018}.
For crystals, calculating invariants is greatly simplified by the  presence of translational symmetry that allows the use of momentum space.
In contrast, topological phases in strongly disordered systems, predicted~\cite{agarwala_topological_2017,mitchell_amorphous_2018, focassio_amorphous_2021, agarwala_higher-order_2020, wang_structural-disorder-induced_2021, Corbae2023} and observed~\cite{mitchell_amorphous_2018,Zhang2023,corbae_evidence_2020}  in e.g., amorphous systems,  require formulating topological invariants in real space.
For topological insulators, multiple options exist.
These are based, for instance, on scattering theory~\cite{Fulga2011, Fulga2012}, non-commutative geometry~\cite{Prodan2010, Prodan2013}, $K$-theory~\cite{Loring2010, Loring2015, Loring2019, Huang2018b,Huang2018, loring_spectral_2020, schulz-baldes_spectral_2021, cerjan_local_2022, cerjan_operator-based_2022}, supercells~\cite{ceresoli_orbital_2007}, Fourier transformation of momentum space formulas~\cite{Kitaev20062, bianco11, marrazzo_locality_2017, Irsigler2019, favata_single-point_2023, Ornellas2022, chen_universal_2022, guzman_geometry_2022}, density matrices~\cite{Hannukainen2022, Hannukainen2023}, symmetry indicators~\cite{marsal_topological_2020, marsal_obstructed_2022}, flattened Hamiltonians~\cite{Jezequel2022}, and effective Hamiltonians in momentum space~\cite{varjas_topological_2019, marsal_topological_2020, spring_amorphous_2021}.

The wide choice of real space topological invariants for insulators contrasts the limited possibilities for diagnosis of topological metals in real space.
The local Chern marker~\cite{bianco11} and the spectral localizer~\cite{Loring2015,Loring2017, Loring2019, Fulga2016,
loring_spectral_2020, Liu2018,schulz-baldes_spectral_2021, Doll2020, Doll2021, Michala2021} have been used even when topological insulator properties coexist with trivial metallic bands~\cite{cerjan_local_2022}, or when the Fermi energy crosses a topological band~\cite{Ceresoli:2006bn}.
However, real space topological invariants capable of detecting unique signatures of topological metallic bands such as Weyl crossings, are scarce.
Topological semimetals that break time-reversal symmetry can be characterized by a non-zero Hall conductivity $\sigma_{xy}\neq 0$~\cite{yang_topological_2019}. 
However, $\sigma_{xy}\neq 0$ can also be a signature of time-reversal broken trivial metals that are e.g., subjected to magnetic fields~\cite{Wang2023W}.
In the presence of time-reversal symmetry, $\sigma_{xy}=0$, rendering the diagnosis even more precarious. 

An important step forward to characterize topological metals in real space was provided in Refs.~\cite{Schulz-Baldes21,Schulz-Baldes2022}.
They suggested that the spectral localizer ${\mathcal{L}}({X},{H})$~\cite{Loring2015, Loring2017, Loring2019} can also diagnose metallic topology.
The spectral localizer ${\mathcal{L}}({X},{H})$ is
a real space operator that quantifies whether the Hamiltonian ${H}$ and the position operator ${X}$ can be continuously deformed to commute without closing the band gap or breaking a symmetry.
Refs.~\cite{Schulz-Baldes21, Schulz-Baldes2022} showed that 
the number of zero-modes of ${\mathcal{L}}$ counts the number of Weyl nodes in topological metals.
Recently this approach was successfully adapted to higher-spin generalizations of Weyl semimetals~\cite{Franca2024}}, known as multifold fermions~\cite{manes_existence_2012,Bradlyn2016,tang_multiple_2017,Chang:2018bb}. 

This body of work shows that the localizer is 
not only successful in identifying topological insulators 
~\cite{Loring2015,Fulga2016,Loring2019,Viesca2019,Loring2019,cerjan_local_2022,cerjan_operator-based_2022} in all Altland-Zirnbauer classes~\cite{altland97}, but also Weyl semimetals.
Crucially, the spectral localizer succeeds in detecting Weyl semimetals even in the presence of time-reversal symmetry.
{However, its properties have not been discussed for trivial metals, and it therefore remains unclear if or how topological and trivial metals differ in what concerns the spectral localizer.}

{In this work we find that the low-lying spectrum of the localizer 
of trivial metals can be inferred from the spectrum of an auxiliary topological insulator
defined in the continuum.}
This remarkable connection becomes transparent once we show that finding 
the zero-modes of the localizer amounts to finding the zero-modes 
of a Dirac equation with a varying mass term.
Such re-branding is advantageous because the localizer spectrum is 
typically computed numerically in a case to case basis
~\cite{Loring2015, Fulga2016, Loring2017, loring_spectral_2020, Liu2018, Loring2019, schulz-baldes_spectral_2021, Doll2020, Doll2021}, while the Dirac equation with varying mass is a recurring problem across physics. 
The analytical solutions to this equation link very distinct phenomena 
including solitons in polyacetyline~\cite{Jackiw1976, Su1979}, boundary 
states of topological insulators~\cite{Qi:2008eu, Karzig2013, Volkov2016, Inhofer2017, Tchoumakov2017}, quantum Hall~\cite{Ludwig1994} and Anderson 
transitions~\cite{Evers2008}, domain walls in high-energy physics~\cite{Rubakov1983, Callan1985} or cosmic string cosmology~\cite{vilenkin2000}, to name a few. 
Here we take advantage of this vast literature to predict the spectral 
differences of ${\mathcal{L}}$ that characterize different types of metals,
 including trivial and topological metals, and gapless chiral edge-states. 
Such understanding allows even to 
go beyond the single-particle picture and conjecture signatures 
of many-body fractional quantum Hall edge states. 
{
\section{Existence of zero-modes of the spectral localizer of metals}
}
{In this section we discuss the motivation and origin of the topological properties of the space of zero-modes of the localizer, both in real and momentum space.
Through the examples we discuss in the next section we showcase how 
this mapping has a main practical advantage: 
the existence and structure of the localizer low-lying states can be calculated without resorting to numerical diagonalization.
}

\bigskip

\subsection{Motivation}
The spectral localizer ${\mathcal{{L}}}(X_0,E_0)$ is a local operator defined 
for a reference energy $E_0$ and position 
$X_0=(x^{(0)}_1,x^{(0)}_2,\cdots,x^{(0)}_d)$, in $d$ 
spatial dimensions~\cite{Loring2015,Loring2017,loring_spectral_2020,Loring2019,schulz-baldes_spectral_2021}
\begin{equation}
\label{eq:locdef}
    {\mathcal{{L}}}(X_0,E_0)=\kappa({X}_j-x^{(0)}_j)\Gamma_j+({H}-E_0)\Gamma_{d+1}.
\end{equation}
Here we assume the Einstein summation convention 
over the spatial index $j=1,2,\cdots d$, ${H}$ is the Hamiltonian in real space, 
and ${X}_j$ are the position operators.
The scalar parameter $\kappa$ fixes the units and the relative weight 
between the two terms.
$\Gamma_j$ are a set of anti-commuting Clifford matrices 
(e.g. the Pauli matrices in $d=2$).
{
For each strong topological insulator class in dimension $d$, 
the operator ${\mathcal{{L}}}(X_0,E_0)$ 
encodes whether ${X}_j$ and ${H}$ can be continued to commuting 
while preserving the necessary symmetries and local gap. 
For example, the Chern number is given by the signature of the localizer, 
the difference between positive and negative eigenvalues
~\cite{Fulga2016, Loring2019, Viesca2019, Loring2019, cerjan_local_2022, cerjan_operator-based_2022}. 
For topological metals, the number of Dirac or Weyl points is equal to the kernel, or number of zero-modes, of the localizer spectrum 
$\sigma[\mathcal{L}]$~\cite{Schulz-Baldes21,Schulz-Baldes2022}.}

{The motivating observation that triggers this work is the so far overlooked} resemblance between Eq.~\eqref{eq:locdef} and a Dirac Hamiltonian
\begin{equation}
\label{eq:Diracdef}
    {H}_{\mathrm{Dirac}}(A_j,m_0)=
    v_F (-i {\partial}_{x_j}-{A}_j(\bs{x}))\Gamma_j + \delta m(\bs{x})\Gamma_{d+1}.
\end{equation}
defined by a Fermi velocity $v_F$, a gauge field $A_j(\bs{x})$ and space-dependent mass $\delta m(\bs{x})=m(\bs{x})-m_0$.
It can be made more explicit by a Fourier transform of Eq.~\eqref{eq:locdef}, 
which can diagonalize the Hamiltonian (${H} \to \epsilon(\bs{k})$) 
and result in the replacement ${X}_j \to i{\partial}_{k_j}$, leading to
\begin{equation}
\label{eq:locdefk}
{\mathcal{{L}}}_k(X_0,E_0)=\kappa(i {\partial}_{k_j}-x^{(0)}_j)\Gamma_j
+(\epsilon(\bs{k})-E_0)\Gamma_{d+1}.
\end{equation}
This operator has now the same form as the Dirac operator 
Eq.~\eqref{eq:Diracdef} if we identify $v_F$ with $\kappa$, $A_j$ 
with $x^{(0)}_j$ and $\epsilon(\bs{k})-E_0$ with $\delta m(\bs{x})$. 
{In doing so we think of the momentum variable $k_j$ in 
Eq.~\eqref{eq:locdefk} playing the role of a space variable $x_j$ 
in the Dirac picture Eq.~\eqref{eq:Diracdef}.}

{This observation is appealing, but not rigorous.
Hence, in the reminder of this section we offer a more rigorous discussion both in the finite Hilbert space where Eq.~\eqref{eq:locdef} is typically defined, and in momentum space of a tight-binding Hamiltonian.
In practice, the Dirac picture turns out to be sufficient to predict the low-lying spectrum of the localizer.}

\bigskip 
{
\subsection{Zero-modes of the localizer in real and momentum space \label{sec:k-space}}
The matrix elements of the localizer, 
written in the basis that diagonalizes the Hamiltonian, read
\begin{equation}
    \label{eq:locdefapp}
        {\mathcal{{L}}}_{mn}(X_0,E_0)=\kappa({X}^{mn}_{j}-x^{(0)}_j\delta_{nm})\Gamma_j
        +({E}_n-E_0)\delta_{nm}\Gamma_{d+1}.
    \end{equation}
where $H\ket{n}=E_n\ket{n}$, ${X}^{mn}_{j}=\bra{m}{X}_{j}\ket{n}$.
For an insulator, i.e. $E_0$ chosen inside the gap, 
the second term is always of the same sign.
However, when the spectrum is dense spectrum, as in a metal, one can always choose $E_0$ such that the second term vanishes for a given $E_n$.
The first term has a zero-mode when $x^{(0)}$ coincides with a site in the system.
These two facts combined show that for a metallic spectrum, we should be able to find a zero-mode of the localizer no matter the topological properties of the metal.

We are thus interested in the space of zero-modes of the localizer, $\mathcal{M}$ defined as
\begin{equation}
    \label{eq:zmspace}
    \mathcal{M}=\set{w \in \mathbb{R}^{d+1}|\sigma[\mathcal{L}(w)] = 0}.
\end{equation}
Here $w$ is a $d+1$ dimensional vector $w=(x^{(0)}_j,E_0)$ and $\sigma(w)$ are the eigenvalues of the localizer for a particular choice of $w$.
Ishiki \textit{et al.}~\cite{Ishiki2018} observed that it is possible to associate a Berry connection and a finite Chern number to $\mathcal{M}$, which is in this mathematical sense, topologically non-trivial.

In this work we wish to take a different, more physical route to bring up the topological content of $\mathcal{M}$, leaving the exact relation between the two approaches for subsequent work. 
To do so, we find the representation of the localizer in the Bloch basis that diagonalizes the Hamiltonian.
This basis is spanned by the eigenstates $\ket{n\bs{k}}$ of band $n$ and momentum $\bs{k}$.
We will make use of a result by 
Blount~\cite{blount1962formalisms}, 
who expressed the position operator in this basis:
\begin{eqnarray}
\nonumber
   \bra{m\bs{k}'}{X}_j\ket{n\bs{k}}&=& \bar{\delta}_{nm} \delta_{\bs{k}\bs{k}'} \mathcal{A}^{j}_{nm}(\bs{k})\\
   \label{eq:posopapp}   
   &+&\delta_{nm} (\delta_{\bs{k}\bs{k}'} \mathcal{A}^{j}_{nm}(\bs{k}) 
   +i \partial_{\bs{k}_j}\delta_{\bs{k}\bs{k}'})
\end{eqnarray}
where $\bar{\delta}_{nm}=(1-\delta_{nm})$, $\delta_{\bs{k}\bs{k}'}= \delta(\bs{k}-\bs{k}')$ and
$\mathcal{A}^{j}_{nm}(\bs{k})= \bra{m\bs{k}} i \partial_{\bs{k}_j}\ket{n\bs{k}}$ is the Berry connection. 
The position operator has both diagonal (${\delta}_{nm}$) and off-diagonal ($\bar{\delta}_{nm}$) contributions which we can use to separate the localizer contributions as well:

\begin{widetext}
    \begin{subnumcases}{\bra{m\bs{k}'}{\mathcal{{L}}}(X_0,E_0)\ket{n\bs{k}}=}
        \kappa( \delta_{\bs{k}\bs{k}'} \mathcal{A}^{j}_{nn}(\bs{k}) + i
 \partial_{\bs{k}_j}\delta_{\bs{k}\bs{k}'}
        -\delta_{\bs{k}\bs{k}'}x^{(0)}_j)\Gamma_j+(\epsilon_n(\bs{k})-E_0)\delta_{\bs{k}\bs{k}'}\Gamma_{d+1}, & $m=n$ \label{eq:meqn}
        \\
        \kappa\delta_{\bs{k}\bs{k}'} \mathcal{A}^{j}_{nm}(\bs{k})\Gamma_j & $n\neq m$ \label{eq:mneqn}
     \end{subnumcases}
\end{widetext}
  
The intraband term $n=m$ expresses the fact that the position operator 
is highly singular in momentum space.
However, observables are well defined, which can be shown
by expressing them in terms of density matrices.
This has thoroughly been discussed 
in the context of non-linear optics, see for example Refs.~\cite{Aversa1995, Sipe2000, Hipolito2016}.
We are interested in trivial metals, which we define by assuming that $E_0$ intersects a single band $n$. 
Hence we focus on the contribution of Eq.~\eqref{eq:meqn} which allows us to write the localizer as
\begin{eqnarray}
\label{eq:locdefappsingle} 
\nonumber
\bra{n\bs{k}}{\mathcal{{L}}}(X_0,E_0)\ket{n\bs{k}}&=& 
\kappa(i \partial_{\bs{k}_j}-x^{(0)}_j+\mathcal{A}^{j}_{nn}(\bs{k}))\Gamma_j\\
&+&(\epsilon_n(\bs{k})-E_0)\Gamma_{d+1}.  
\end{eqnarray}
Eq.~\eqref{eq:locdefappsingle} has the form of a Dirac particle coupled to a vector potential and a mass term:
\begin{eqnarray}
\label{eq:locdefappsingledirac} 
\nonumber
{H}_{\mathrm{Dirac}}(A_j,m_0)&=&v_F (-i {\partial}_{x_j}-{A}_j(\bs{x}))\Gamma_j\\
&+&(m(\bs{x})-m_0)\Gamma_{d+1}. 
\end{eqnarray}
if we identify $i \partial_{k_j} \leftrightarrow - i\partial_{x_j}$, $v_F\leftrightarrow\kappa$, 
$A_j(\bs{x})\leftrightarrow x^{(0)}_j-\mathcal{A}^{j}_{nn}(\bs{k})$ 
and $\epsilon(\bs{k})-E_0\leftrightarrow m(\bs{x})-m_0=\delta m(\bs{x})$.  
There is an emergent gauge degree of freedom carried by the 
Berry connection $\mathcal{A}^{j}_{nn}(\bs{k})$. 
However, since the low-lying spectrum of the localizer is only sensitive to a 
local region in $\bs{k}$ space, close to where $\epsilon(\mathbf{k})-E_0=0$, 
the gauge connection $\mathcal{A}^{j}_{nn}(\bs{k})$ can be gauged away
\footnote{We are indebted to Prof. Schulz-Baldes for this remark. 
This was rigorously proven in Ref.~\cite{Schulz-Baldes2022} for the case of a Weyl semimetal}.
With this simplification, we arrive to Eq.~\eqref{eq:locdefk}.

When reasoning with the Dirac equation, we will be assuming the continuum 
limit approximation, while numerical calculations in real space require a finite Hilbert space.  
As we will see, both will lead to consistent results, showcasing the usefulness of 
the long-wavelength Dirac picture.

In the momentum space picture, zero-modes of the localizer amount 
to zero-modes of a Dirac equation.
While this is a mere recasting of the localizer, this viewpoint turns out to be advantageous.
One can now take advantage of literature on the physical consequences and topological properties of zero-modes of the Dirac equation, to infer properties of the localizer zero-mode space without the need of exactly diagonalizing the operator.

We now present a set of benchmarks that compare the predictions made using this momentum space construction to exact diagonalization of the localizer in real space.
}

\section{Benchmarks}

{Here we give numerical and analytical support to the general considerations above by studying several examples.
We focus specifically on showing the differences between trivial and topological metals.
We conclude with the example of a fractional quantum Hall edge.}

\subsection{$d=1$ wire}
\begin{figure}
    \centering
    \includegraphics[width=\columnwidth]{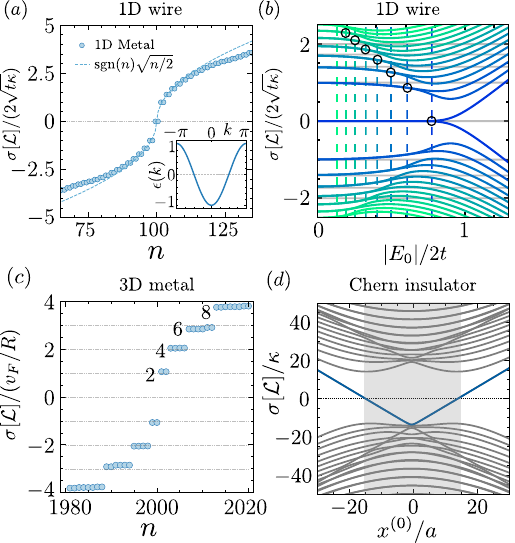}
    \caption{\label{fig:1D3Dwire} 
    Localizer spectrum $\sigma[\mathcal{L}]$ of bulk and edge metals. 
    (a) $\sigma[\mathcal{L}]$ (circles) for a 1D finite chain of length $L=100$ at $E_0=0$ in units of $2\sqrt{t\kappa}$ as a function of 
    eigenvalue index $n$, for $\kappa=0.05$, and $t=1/2$. The low-lying spectrum is doubly degenerate, and follows a $\mathrm{sgn}(n)\sqrt{n/2}$ law (dashed line). 
    This is the characteristic $\sqrt{n}$ law for relativistic fermions, with the $1/2$ accounting for the double degeneracy. Inset: 1D dispersion relation $\epsilon(k)$. (b) 
    Same spectrum as in (a) now as a function of $E_0$. As $E_0$ increases, the two-fold degeneracy is lifted where the dashed vertical lines $E_0\approx 2t-l_\mathrm{osc}\sqrt{2n+1}$, 
    with $|n|=0,1,2,\cdots$, meet the localizer levels (black circles). Horizontal gray lines indicate  the pseudo-relativistic $\sqrt{n}$ law. (c) {$\sigma[\mathcal{L}]$ 
    (in units of $v_F/R=\kappa/\sqrt{(6t+E_0)/t}$) for a 3D finite cube of linear length $L=10$ at $E_0=-2.5$, with $\kappa=0.1$, $t=1/2$. The low lying states are spaced 
    in multiples of $v_F/R$ (dashed lines) and their degeneracy, given by the sequence $2,4,6,8,10,\cdots$, is equal to that of a finite spherical topological insulator of 
    radius $R$, see Eq.~\eqref{eq:TIsphere}.} (d) 
    $\sigma[\mathcal{L}]$ as a function of $X_0=(x^{(0)},0.)$ for a square Chern insulator of linear dimension $L=30$, with $\kappa=0.01$ and $M/t=-1$. 
    In the Dirac picture, we expect a zero at the boundary, $|x_0|=L/2=15$, as observed numerically. The shaded region indicates the system's interior, 
    where the localizer signature and Chern number equal $-1$~\cite{Loring2015,Loring2019,Viesca2019}.}
\end{figure}

%
As a warm up exercise we wish to find how the localizer spectrum changes as we interpolate between a one-dimensional (1D) trivial parabolic dispersion and a 1D Weyl Hamiltonian.
This can be achieved by studying the 1D tight-binding lattice Hamiltonian
\begin{equation}
\label{eq:1Dwire}
    H= -t \sum_{n=1}^L c^{\dagger}_n c_{n+1}^{}+\mathrm{h.c.},
\end{equation}
where we assume the lattice constant equals $a =1$, $c^{\dagger}_n$ ($c^{}_n)$ is the creation (annihilation) operator of a particle at site $n$, and $L$ is the length of the chain.
This chain has a bulk energy spectrum $\epsilon(k)=-2t\cos(k)$, see inset of Fig.~\ref{fig:1D3Dwire}(a). 
Close to the bottom of the band, at $k=0$, the dispersion relation is  $\epsilon(k)\approx-2t+tk^2+\cdots$.
We take this as a definition of a 1D trivial metal: a single-band, parabolic dispersion at momentum $k\sim0$ and energy $\epsilon(k)\sim-2t$ with effective mass $m=\frac{1}{2t}$.
For energies $E_0\approx 0$, we expand around $k=\pm\pi/2$ and obtain $\epsilon(k)_{\pm}\approx - 2t (\pm k + \frac{\pi}{2})$ to linear order .
The left and right moving dispersion relation can be compactly encoded in a 1D Weyl Hamiltonian of the form $H_{W}=-2t (k\tau_z+\frac{\pi}{2}\tau_0)$, 
where $\tau_z$ and $\tau_0$ are the third Pauli matrix and the $2\times 2$ identity matrix, respectively. 
As we will see the number of zero-energy modes of the spectral localizer~\cite{Schulz-Baldes21, Schulz-Baldes2022} is not enough to distinguish the 1D parabolic and Weyl limits of Eq.~\eqref{eq:1Dwire}. 
Mapping to a Dirac Hamiltonian we will be able to predict the differences in the localizer spectrum between these two limits.

At energies close to $E_0 =0$, $\epsilon(k)$ is linear, and we may write the localizer in momentum space as 
\begin{equation}
\label{eq:locdefmom2}
    {\mathcal{{L}}}_k(0,E_0)_\pm=i \kappa \partial_{k}\sigma_x - 2t (\pm k + \frac{\pi}{2})\sigma_z.
\end{equation}
where we chose $\Gamma_1=\sigma_x$ and $\Gamma_{2}=\sigma_z$.
Interpreting $k$ as a coordinate, as discussed below Eq.~\eqref{eq:locdefk}, we recognize this operator as the 1D Dirac Hamiltonian
\begin{equation}
\label{eq:dirac1d}
    H_{\mathrm{Dirac}}^{\pm}=-i v_F\partial_{x}\sigma_x+\delta m_{\pm}(x)\sigma_z
\end{equation}
with $v_F=\kappa$ and a linearly varying mass $\delta m(x) = -2t(\pm x + \frac{\pi}{2})$, of which the solutions are well known (see e.g. Ref.~\cite{Karzig2013})
For each chirality ($\pm$), finding its spectrum and eigenfunctions analytically amounts to finding that of a particle in a harmonic oscillator~\cite{Karzig2013}, with a characteristic length scale $l_{\mathrm{osc}}= \sqrt{\kappa/2t}$.
The localizer has thus a spectrum $\sigma_n=\mathrm{sgn}(n)2\sqrt{t\kappa n}$.
We notice that not only the zero-mode is doubly degenerate, as found in Ref.~\onlinecite{Schulz-Baldes21}, but also that all states show this degeneracy, provided that $\epsilon(k)$ can be approximated to be linear.
The spectrum displays $(\sigma,-\sigma)$ doublets because of the chiral symmetry $\mathcal{C} = \sigma_y$ of the localizer that imposes $\mathcal{C} \mathcal{L} \mathcal{C}^{\dagger} = -\mathcal{L}$.

Numerically diagonalizing the 1D localizer Eq.~\eqref{eq:locdef} with $\Gamma_1=\sigma_x$ and $\Gamma_{2}=\sigma_z$ confirms our analytical expectations.
Choosing $x^{(0)}$ at the center of the system, and $H$ to be Eq.~\eqref{eq:1Dwire} we obtain the spectrum shown in Fig.~\ref{fig:1D3Dwire}, 
shown as a function of eigenvalue index $n$ in (a) and $E_0$ (b).
For $E_0=0$, the low energy spectrum consists of doubly degenerate $(\sigma,-\sigma)$ pairs.
As predicted, the low-lying spectrum $\sigma$ follows a $\sqrt{n}$ dependence, up to an energy determined by the breakdown of the 
linear approximation to $\epsilon(k)$ leading to Eq.~\eqref{eq:dirac1d}, see Fig.~\ref{fig:1D3Dwire}(a).

As $|E_0|$ approaches $2t$, i.e., the trivial metal limit, we observe in Fig.~\ref{fig:1D3Dwire}(b) that the localizer spectrum looses its double degeneracy.
Higher-energy states loose their degeneracy farther away from $|E_0|=2t$, compared to the zero-mode, which looses its degeneracy close to $|E_0| \approx 2t$.
The fact that the zero-mode remains doubly degenerate can be predicted by writing the localizer in momentum space close to $E_0\approx -2t$:
\begin{equation}
\label{eq:locdefmom3}
{\mathcal{{L}}}_k(0,E_0)= i\kappa\partial_{k}\sigma_x+(-2t+tk^2-E_0)\sigma_z.
\end{equation}
Interpreting $k$ as a coordinate we recognize once more the 1D Dirac Hamiltonian with $v_F=\kappa$. 
The mass varies parabolically $\delta m(x) = tx^2-(E_0+2t)$, and not linearly as in Eq.~\eqref{eq:locdefmom2}.
If $|E_0|< 2t$ such mass profile crosses zero twice, at $\pm \sqrt{(2t+E_0)/t}$, and thus we expect two degenerate zero-modes, as confirmed numerically.

We still require to explain the difference between the localizer spectrum at $|E_0|\approx 2t$ (trivial metal)  and at $E_0 \approx0$ (Weyl metal), 
in particular that higher energy levels start dispersing and are no longer doubly degenerate as $E_0$ is increased.
These two properties are understood by drawing a parallelism to Dirac Landau levels, which start dispersing and loose their degeneracy as they approach 
an edge~\cite{Brey:2006gx, Coissard2022}. 
Here reaching the bandwidth as we increase $E_0$ acts like a sample edge; the point where the Landau levels start dispersing and loose their degeneracy 
is determined by their average extent, $\braket{r_n} = l_\mathrm{osc}\sqrt{2n+1}$~\cite{Coissard2022}.
This prediction matches well with the numerical diagonalization, as marked by the vertical dashed lines in Fig.~\ref{fig:1D3Dwire}(b).
In the Dirac equation language, the two zero-mode solutions begin to hybridize when $|E_0|\approx 2t$.

The eigenstates of the localizer also confirm the Dirac interpretation.
For a single-band Hamiltonian with band dispersion $\epsilon(k)-E_0$, the wave-function of the localizer zero-modes are of the form:
\begin{equation}
\label{eq:psi0}
    \psi_{\mathrm{zm}}(k) = N \mathrm{exp}\left(\pm\dfrac{1}{\kappa}\int^{k}dk'(\epsilon(k')-E_0)\right)\psi^{\pm}_0,
\end{equation}
where $\psi^{\pm}_0=(1,\pm1)^T$ are two chiral eigenstates and $N$ is a normalization constant.
These are the well known (Jackiw-Rebbi) solutions of the Dirac Hamiltonian with a varying mass term~\cite{Jackiw1976,Qi:2008eu,Ludwig1994}, provided we interpret $k$ as a coordinate variable.
There are two normalizable solutions, localized close to where the mass crosses $E_0$.
We have confirmed this expectation numerically by projecting the localizer zero-energy modes at $E_0=0$, obtained by exact diagonalization, 
onto plane waves $\mathrm{exp}(i k x)$.
The zero-modes have the spinor structure set by Eq.~\eqref{eq:psi0}, and are localized at $k=\pm\pi/2$, where the dispersion relation crosses $E_0$, 
i.e. where the Dirac mass changes sign. 
Lastly, for energies $E_0>2t$, the localizer spectrum is gapped (see Fig.~\ref{fig:1D3Dwire}(b)).
This is expected in the Dirac picture because the Dirac mass is always positive if $E_0>2t$, implying no sign change and no zero-modes.
From Eq.~\eqref{eq:locdefmom3} we see that the mass $\delta m$ takes its most negative value at $k=0$, $\delta m(0) =-(2t+E_0)$. 
This value constrains how much the mass can vary as we change $k$ in Eq.~\eqref{eq:psi0}, and thus how localized are the eigenmodes. 
At $E_0\geq 2t$ the zero-modes hybridize and  Eq.~\eqref{eq:psi0} is no longer a solution, resulting in a gapped spectrum.
We reach an analogous conclusion for $E_0<-2t$, provided we expand the localizer around $E_0\approx2t$ and momentum $k=\pi$.

In short mapping the localizer to a Dirac Hamiltonian allows us to map the problem of finding its spectrum to solving a Dirac Hamiltonian with a varying mass. 
One may ask if and how these considerations carry over to three-dimensions (3D).
We address this point next.

\subsection{$d=3$ Trivial metal}

The distinction between a trivial and a topological metal becomes explicit in 3D. 
Let's first predict, using the Dirac picture, the localizer spectrum of a trivial metal.
To define a trivial metal in 3D we generalize the 1D hamiltonian Eq.~\eqref{eq:1Dwire} to
\begin{equation}
\label{eq:3Dwire}
    H= -t \sum_{\mathbf{x}} \sum^{3}_{j=1} c^{\dagger}_{\mathbf{x}} c_{\mathbf{x}+\hat{x}_j}^{}+\mathrm{h.c.},
\end{equation}
where $c^{\dagger}_{\mathbf{x}}$ ($c^{}_{\mathbf{x}}$) is the creation (annihilation) operator of a particle at site $\mathbf{x}$, and $\hat{x}_j$ is a unit vector in the $j$-th direction. 
The Hamiltonian Eq.~\eqref{eq:3Dwire} is defined in a cubic lattice, and has a bulk energy spectrum $\epsilon(\bs{k})=-2t\sum_j\cos(k_j)$ .
The parabolic dispersion near the band minimum is well captured by the expansion close to $\bs{k}=0$ such that $\epsilon(\bs{k})\approx-6t+t|\bs{k}|^2+\cdots$. 
For low fillings, this dispersion relation produces a nearly spherical Fermi surface found in good metals~\cite{Ashcroft76}.
We therefore study the spectral localizer at low fillings, i.e., we set $E_0$ close to the band edge.

According to our Dirac Hamiltonian picture the varying mass is determined by the parabolic dispersion relation close to $|E_0|\approx -6t$. 
Hence, the Dirac mass is negative close to $|\bs{k}|=0$, and grows parabolically to be positive at large $|\bs{k}|$.
The Dirac mass hence vanishes at the sphere defined by $\epsilon(\bs{k})=E_0 =-6t+t|\bs{k}|^2$, which defines the boundary between two 3D time-reversal invariant insulators with masses of opposite signs. 
Hence at such boundary, we expect a gapless boundary state that has a low-lying spectrum determined by a two-dimensional (2D) spherical Dirac equation.
Such low-lying spectrum was found in Ref.~\onlinecite{Imura2012} and is given by
\begin{equation}
\label{eq:TIsphere}
    \varepsilon_{l,m}=\pm\dfrac{v_F}{R}|l+|m|+1/2|=\pm(1,2,3,4,\cdots)\dfrac{v_F}{R},
\end{equation}
where $R$ is the radius of the topological insulator, {or equivalently the Fermi surface sphere in our case}. 
The indices $l=0,1,2,\cdots$ and $m=\pm \frac{1}{2}, \pm\frac{3}{2},\cdots$ determine degeneracies of the energy levels to be $2,4,6,8,10,\cdots$.
As before $v_F=\kappa$ is the velocity of the effective Dirac Hamiltonian.
{When $R\to\infty$ the spectrum becomes gapless, as expected for a topological insulator in the thermodynamic limit.}
{In our case, we need to keep the radius of the effective topological insulator finite, since it} is given by the equation $R^2=(6t+E_0)/t$.
With $t=1/2$ we can set $R=1$ by choosing $E_0=-2.5$.
Solving for the spectral localizer in 3D with $\kappa=0.1$, $\Gamma_{j}=\sigma_j\otimes \sigma_z$, $\Gamma_4=\sigma_0\otimes \sigma_x$ and plotting the energies in units of $v_F/R$ results in Fig.~\ref{fig:1D3Dwire}(c).
The degeneracies of states, the spacing between them and their energies match perfectly with the expectation for a surface state of a spherical topological insulator, see Eq.~\eqref{eq:TIsphere}.

Once more the knowledge of the solution to the Dirac Hamiltonian allowed us to predict the low-lying spectrum of the localizer. 
For both 1D and 3D, finding the localizer spectrum of a trivial metal amounts to finding the surface state spectrum of a finite topological insulator in the continuum limit. 
In the following, we show how the results of Refs.~\cite{Schulz-Baldes2022, Schulz-Baldes21} and the results here can be used to distinguish trivial and topological semimetals using the spectrum of the localizer.

\begin{figure}
    \centering
    \includegraphics[width=\columnwidth]{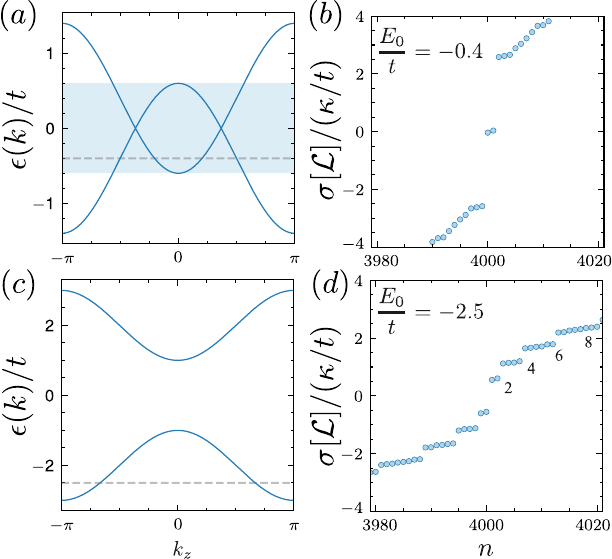}
    \caption{\label{fig:3DWSM}
    Localizer spectrum $\sigma[\mathcal{L}]$ of a Weyl semimetal model in different limits. (a) Band structure of $H_{\mathrm{WSM}}$ 
    for $M/t=2.4$ displaying Weyl nodes at $\epsilon/t=0$. The shaded area spans the energy region where the Weyl bands are defined. (b) 
    Corresponding localizer spectrum for $E_0/t=-0.4$ (dashed line in (a)) with $\kappa=0.1$ for a 3D finite cube of linear length $L=10$. 
    The two mid-gap modes are present whenever $E_0/t$ lies within the shaded region in (a). (c) Band structure in the trivial insulator 
    limit separated Weyl nodes ($M/t=4$). (d) Corresponding localizer spectrum for $E_0/t=-2.5$ (dashed line in (c))  for $\kappa=0.1$, 
    for a 3D finite cube of linear length $L=10$. The spectrum and degeneracy equal those of trivial metal, i.e. those of a finite spherical 
    topological insulator, given respectively by Eq.~\eqref{eq:TIsphere} (dashed lines) and the sequence $2,4,6,8,\cdots$.
    }
\end{figure}

\subsection{$d=3$ Weyl semi-metal}
To interpolate between a trivial metal and a Weyl semimetal we use the two-band Hamiltonian
\begin{eqnarray}
\label{eq:WSM}
    &&H_{\mathrm{WSM}}= H_{CI}-t\cos(k_z)\tau_z,\\
\label{eq:CI}
&&
\begin{split}
    H_{\mathrm{CI}}=& -t [\sin(k_x)\tau_x+\sin(k_y)\tau_y+ \\
    & \cos(k_x)\tau_z+\cos(k_y)\tau_z ]+M\tau_z,
\end{split}
\end{eqnarray}
written in momentum space in terms of a Chern insulator Hamiltonian $H_{\mathrm{CI}}$~\cite{QiWuZhang2006}. For $t<M/t<3t$, $H_{\mathrm{WSM}}$ 
displays two Weyl points at zero energy and $\bs{k}=(0,0,\pm k_W)$ with $k_W=\arccos(M/t-2)$, see Fig.~\ref{fig:3DWSM}(a). 
For $|M/t|>3t$, $H_{\mathrm{WSM}}$ has two trivial parabolic bands separated by a gap, see Fig.~\ref{fig:3DWSM}(c).

In the Weyl phase, solving numerically for the spectrum of the localizer at $E_0=0$ in the bulk of the system  
(with $\Gamma_{j}=\sigma_j\otimes \sigma_z$, $\Gamma_4=\sigma_0\otimes \sigma_x$) we observe two mid-gap states, predicted in Refs.~\cite{Schulz-Baldes2022, Schulz-Baldes21}. 
For $E_0/t\neq 0$, we observe that the mid-gap states remain (see Fig.~\ref{fig:3DWSM}(b)) so long as $E_0/t$ crosses
 the two bands forming the Weyl cones (shaded region in Fig.~\ref{fig:3DWSM}(a)).
Additionally, the states are not exactly degenerate; their separation increases as the Weyl nodes come closer in momentum space.

Both of these features are explained with our interpretation of the localizer as a  Dirac Hamiltonian with a varying mass. 
First, because the Weyl nodes map to zero-modes of the Dirac Hamiltonian, the closer they are in momentum space, the larger their overlap, and finite size gap.
Second, when $E_0/t$ crosses one single band, either because it goes well beyond the shaded area or because the Weyl cones are absent 
(as in Fig.~\ref{fig:3DWSM}(c)), we recover the localizer spectrum of a trivial 3D metal (Fig.~\ref{fig:3DWSM}(d)).
This is confirmed by the degeneracy counting of the low lying localizer spectrum, and the low lying states being approximately 
equally spaced of order $\kappa/t$, which coincide with Eq.~\eqref{eq:TIsphere}, see Fig.~\ref{fig:3DWSM}(d). 
Further supporting evidence based on how the localizer spectrum evolves for different parameters is presented in Appendix \ref{app:3DWey}.

With the above analysis, we have learned how to distinguish a Weyl semimetal from a trivial metal using the localizer. 
In 1D  the Weyl and a parabolic dispersion differ in the degeneracy of the localizer spectrum. 
In 3D the spectral differences between the trivial and topological metal show both in degeneracy and energetics of the low lying states. 
These properties of the localizer are tied to those of the Dirac Hamiltonian with a varying mass. 

\subsection{Spectral localizer of 1D chiral Luttinger liquids}

To finish, we rephrase a known result for the localizer of a Chern insulator using our Dirac picture.
This will allow us to conjecture how many-body fractional quantum Hall edges might emerge in the spectrum of the localizer, which remains an open problem~\cite{cerjan_local_2022}.

Chiral gapless states occur at the boundaries of the quantum Hall effect and Chern insulators. 
The signature of the localizer, the difference between the number of positive and negative eigenvalues, can distinguish between 2D Chern and trivial insulator phases~\cite{Loring2015, Loring2019, Viesca2019, cerjan_local_2022, cerjan_operator-based_2022}.
The difference between the two are eigenstates that cross as we move $x^{(0)}_j$ from far outside the system ($x^{(0)}\to \infty$),
where the spectrum is particle hole symmetric, to inside the system ($x^{(0)}=0$).
We reproduce this expectation in Fig.~\ref{fig:1D3Dwire}(d) for $H_{CI}$ in Eq.~\eqref{eq:CI}. 
The lowest lying state crosses zero as a function of $x^{(0)}_j$, changing the localizer's signature from zero outside to one 
inside of the Chern insulator (shaded region), see Fig.~\ref{fig:1D3Dwire}(d). 

The zero-mode can be predicted using the varying mass Dirac picture.
Recall that the dispersion of a chiral edge state is $\epsilon_{k}=+v_F k$, where $k$ is the momentum parallel to the edge. 
Hence, when $x^{(0)}_j$ is close to the edge and such a dispersion is a good description of the otherwise gapped Hamiltonian, 
the localizer takes once more the form Eq.~\eqref{eq:locdefmom2}. 
From our previous discussion, we expect a zero-mode at each edge, in agreement with Fig.~\ref{fig:1D3Dwire}(d).
The advantage of the Dirac picture is that we can now conjecture what would be the spectral localizer signature of an edge mode of a 
fractional quantum Hall edge state, which has not been previously discussed to our knowledge.
For example, close to the edge of a Laughlin fractional quantum Hall state at filling fraction $\nu=1/m$, the edge Hamiltonian is that of a chiral boson, dispersing as $\epsilon_{k}=+v_F k$. 
The Fermi velocity is a non-universal factor that depends on $\nu$ and residual interactions~\cite{giamarchi2004quantum}. 
Using our Dirac picture, we predict one zero-mode of the localizer at each edge, as in the non-interacting case.
This prediction may be confirmed numerically in lattice models, using for example finite density matrix renormalization group calculations (see e.g. Ref.~\cite{Dong2018}).

{

\section{Relevance to experiments}

The localizer is especially valuable to describe photonic crystals~\cite{cerjan_local_2022, cerjan_operator-based_2022}, where the tight-binding approximation breaks down at large wavelengths.
These works put forward tools to establish topology in photonic topological gaps.
The localizer succeeds in detecting topology even when topological surface or edge states hybridize with trivial metallic bands~\cite{cerjan_local_2022}.

Our results make it possible to distinguish topological and trivial  gapless photonic systems, as those realized recently in Ref.~\cite{Jorg2022}, by comparing their localizer spectrum.
We foresee the ideas we discuss here can be useful to determine the manifold of zero-modes of the spectral localizer that describe metals defined in non-orientable manifolds, as recently realized in experiment~\cite{Grossi23}. 
We leave this avenue for future work.

Lastly, we have applied the localizer to detect amorphous topological metals in Ref.~\cite{Franca2024}. 
A tight-binding model of crystalline Cobalt Silicide shows zero-modes of the localizer separated by a gap from a continuum of states, as in Fig.~\ref{fig:3DWSM}(b).
These are due to the existence of multifold fermions~\cite{manes_existence_2012,Bradlyn2016,tang_multiple_2017,Chang:2018bb}, higher-spin generalizations of Weyl fermions.
Adding disorder lifts the zero-modes closing the gap and eventually leading to a trivial Anderson insulator.
}

\section{Conclusions}
{

In this work we have characterized the low-lying spectrum of the spectral localizer for trivial and topological metals.
We have shown that it is always possible to choose a reference energy $E_0$ of the spectral localizer that results in zero-modes of the localizer.
We have also shown that, although zero-modes of the spectral localizer are not unique to Weyl semimetals, the spectral localizer spectrum is different for trivial and topological metals.
Our approach is based on the similarity between the spectral localizer in momentum space and the Dirac Hamiltonian with a varying mass. 
This resemblance allows us to infer analytically the low-energy spectrum of the spectral localizer.

Numerical and analytical results coincided in showcasing the difference between localizer's zero-modes for topological and trivial metals.
Our results confirm that topological metals have a number of zero-modes that equals the number of Weyl nodes,
separated by a gap to a continuum of states~\cite{schulz-baldes_spectral_2021, Schulz-Baldes2022}. 
This signature contrasts that of trivial metals which have a low lying spectrum that 
coincides with that of the boundary states of an auxiliary 
topological insulator, defined by the band dispersion of the metallic band.
For these systems, the localizer spectrum becomes continuous in the limit of large Fermi momentum, in contrast to the spectral localizer spectrum for Weyl semimetals that retains a gap between the zero-modes and a continuum of states.
As shown in Ref.~\cite{Franca2024}, this gap encodes the topological robustness of Weyl semimetals and higher-spin generalizations to disorder.

As an outlook, the understanding of the localizer for metallic bands that we put forward
hinges on the spectrum of topological insulators surface states.
This connection suggests that the localizer could be leveraged
as a tool to classify different types of metals, a subject we leave for future work.
Moreover, the localizer is suitable to study disorder systems, 
including amorphous~\cite{Corbae2023,corbae_evidence_2020,Franca2024}, 
polycrystals or quasicrystals\cite{Zilberberg:21}, as it is defined in real space. 
The Dirac Hamiltonian picture we presented can be particularly useful to map 
the disordered topological metal, a system posing 
fundamental open questions~\cite{Pixley2021}, 
to a disordered Dirac equation, which has been 
thoroughly studied in the literature~\cite{Ludwig1994}.

Lastly, the localizer is well defined for interacting systems.  
Here, we used the Dirac picture to conjecture the edge fingerprint of a 
fractional quantum Hall edge. We expect that the localizer will be also 
useful to distinguish other interesting groundstates such spin-liquids in real space, 
even in situations where disorder is dominant~\cite{Cassella2022, Grushin2022}.
}

\section*{Acknowledgements} 

We thank H.~Schulz-Baldes for inspiring conversations that started this work, A. Cerjan, T. Loring for useful comments on the manuscript, 
and J. H. Bardarson, C. Repellin and O. Pozo for discussions. A.G.G. and S. F. acknowledge financial support from the European Union Horizon 
2020 research and innovation program under grant agreement No. 829044 (SCHINES).
A. G. G. is also supported by the European Research Council (ERC) Consolidator grant under grant agreement No. 101042707 (TOPOMORPH). 
The code used to generate our results is freely accessible 
through Zenodo in Ref.~\cite{franca_selma_2023_7795965}

%

\clearpage
\newpage

\setcounter{secnumdepth}{5}
\renewcommand{\theparagraph}{\bf \thesubsubsection.\arabic{paragraph}}

\renewcommand{\thefigure}{S\arabic{figure}}
\setcounter{figure}{0} 

\appendix

\onecolumngrid

\section{\label{app:3DWey}Further discussion of the spectral localizer of a 3D two-band Weyl semimetal model}

\begin{figure*}
    \centering
    \includegraphics[width=0.8\linewidth]{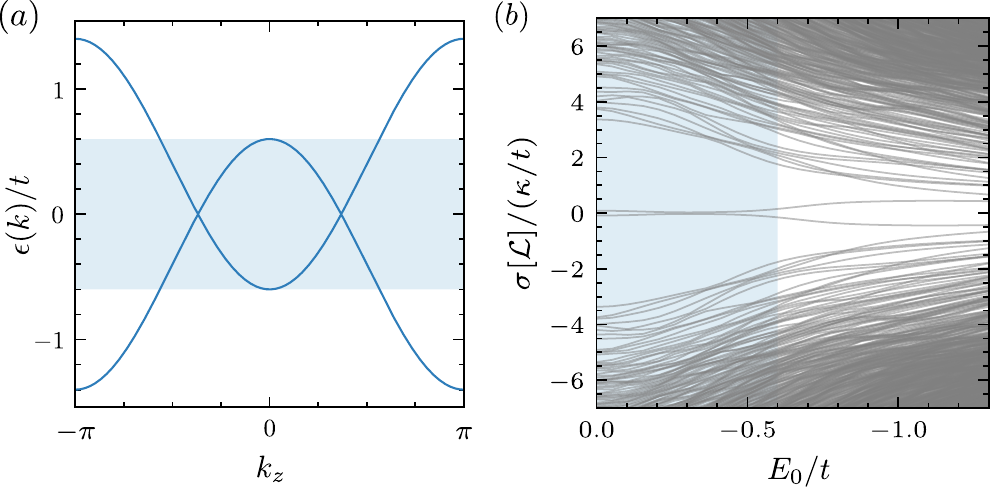}
    \caption{\label{fig:3DmetalLocevolapp} Evolution of the spectral localizer as a function of $E_0/t$. (a) Band structure of the 3D two-band 
    WSM phase with $M/t=2.4$, as in Fig.~\ref{fig:3DWSM}(a) of the main text. (b) Corresponding localizer spectrum as a function of $E_0/t$.
     The two-fold mid-gap states are pinned close to zero energy only when $E_0/t$ lies within the blue shaded region where the Weyl spectrum is defined.
    }
\end{figure*}

In the main text we showed that localizer spectra of the Hamiltonian
\begin{equation}
    H_{\mathrm{WSM}} = -t\sin(k_x)\tau_x-t\sin(k_y)\tau_y+(M-t\sum\limits_{j={x,y,z}}\cos(k_j))\tau_z
\end{equation}
can interpolate between the signatures of Weyl and trivial metals. 
Here we are interested in exploring how the localizer spectra, $\sigma[\mathcal{L}]$, changes as we vary the Weyl node separation and $E_0$.

First, in Fig.~\ref{fig:3DmetalLocevolapp} we represent the band structure and localizer spectrum as a function of $E_0/t$ for the phase 
discussed in the main text in Fig.~\ref{fig:3DWSM}(a) and (b).
As advertised in the main text, the two mid-gap states signaling the two Weyl cones exist so long as $E_0/t$ is smaller than the van Hove energy
 which conects the two nodes (blue shaded area in Fig.~\ref{fig:3DmetalLocevolapp}(a,b)).

\begin{figure*}
    \centering
    \includegraphics[width=\linewidth]{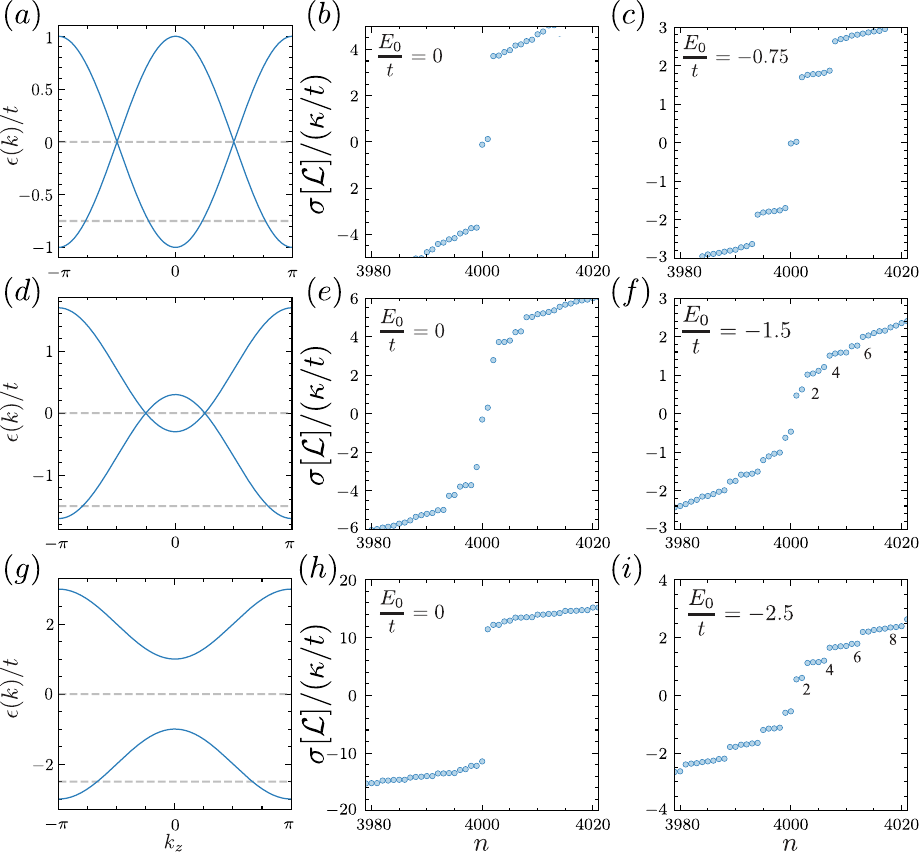}
    \caption{\label{fig:3DmetalLocapp} 
    Evolution of the localizer spectrum for different band structures. (a,d,g) show the band structure for a 3D two-band WSM model with $M/t=2,2.7,4$, respectively. 
    Dashed lines show the values of $E_0/t$ for which we compute the localizer spectrum $\sigma[\mathcal{L}]$ that is plotted in units of $\kappa/t$. (b,c) show $\sigma[\mathcal{L}]$ calculated for a band structure shown in (a) with maximally separated Weyl cones, obtained for $E_0/t=0$ and $E_0/t=-0.75$, respectively, with $\kappa=0.1$. 
    In both panels, we see two mid-gap states corresponding to the two Weyl cones. 
    (e,f) show $\sigma[\mathcal{L}]$ calculated for a band structure shown in (d), obtained for $E_0/t=0$ and $E_0/t=-1.5$, respectively, with $\kappa=0.1$. 
    The first panel shows two mid-gap states corresponding to the two Weyl 
    cones while the second panel begins to display the characteristic spectrum of a trivial metal. The degeneracy is dictated by the sequence of a spherical topological surface state, as discussed in the main text.  
    (h,i) show $\sigma[\mathcal{L}]$ corresponding to a two-band trivial insulator with a band structure shown in (c), 
    at $E_0/t=0$ and $E_0/t=-2.5$, respectively, with $\kappa=0.1$. 
    The first panel shows a clear gap  while the second panel has a well developed spectrum characteristic of a trivial metal. 
    The multiplets are equally spaced and their degeneracy is dictated by the sequence of a spherical topological surface state, as discussed in the main text.
    }
\end{figure*}

To gain more insight on the evolution of the localizer spectrum as a function of $M/t$ and $E_0/t$ we now focus on three particular cases, displayed in 
Fig. \ref{fig:3DmetalLocapp}.
Focusing on the first row, Fig. \ref{fig:3DmetalLocapp}(a) shows a Weyl semimetal with maximum Weyl node separation $\Delta K_W=\pi$, obtained by setting $M/t=2$.
For $E_0/t=0$, we recover a gapped spectrum with two zero-modes, see Fig.~\ref{fig:3DmetalLocapp}(b), as predicted by Refs.~\cite{Schulz-Baldes2022,Schulz-Baldes21}. 
As seen from Fig.~\ref{fig:3DmetalLocapp}(c), we observe that lowering $E_0/t$ does not change this fact, so long as $E_0/t$ crosses the two bands composing the Weyl cones.

For the second row, we choose $M/t=2.7$ and the resulting band structure is shown in Fig.~\ref{fig:3DmetalLocapp}(d).
Here the Weyl node separation is smaller than in Fig.~\ref{fig:3DmetalLocapp}(a).
At $E_0/t=0$ (Fig.~\ref{fig:3DmetalLocapp}(e)) we again find two zero-modes corresponding to the Weyl states, albeit separated in energy. 
Comparing Figs.~\ref{fig:3DmetalLocapp}(b) and (e), we observe that larger Weyl node separations induce a smaller splitting of the mid-gap localizer states.
As discussed in the main text, this agrees with the intuition based on the Dirac Hamiltonian picture: the two Dirac zero-modes of the localizer 
are closer together as the Weyl nodes become closer in momentum space, allowing for the two zero-modes to hybridize.
The two zero-modes lie within the gap until we lower $E_0$ below the van Hove energy, as seen previously in Fig.~\ref{fig:3DmetalLocevolapp}(b).
At energies $E_0/t=-1.5$ the spectrum is dominated by a single trivial parabolic band.
Hence, the localizer spectrum shown in Fig.~\ref{fig:3DmetalLocapp}(f) displays energies and degeneracies that coincide with the surface spectrum of a 3D spherical topological insulator that is discussed in the main text.

Lastly, in the third row, Fig.~\ref{fig:3DmetalLocapp} (c) shows the band structure when $M/t=4$, such that the spectrum is gapped at half-filling. 
When $E_0$ lies within the gap the spectrum of the localizer is also gapped.
However, as we decrease $E_0$ to cross the band, we recover a trivial metal, signaled once more by the energies and degeneracies of a 3D spherical topological insulator.

\end{document}